\documentclass[10pt]{iopart}

\usepackage{graphicx}
\usepackage{dcolumn}
\usepackage{bm}
\usepackage{siunitx}
\usepackage{xcolor}
\usepackage{float}
\usepackage[utf8x]{inputenc}
\usepackage{hyperref}

\begin{document}


\title{Highly symmetric and tunable tunnel couplings in InAs/InP nanowire heterostructure quantum dots}



\author{Frederick S. Thomas$^{*,1}$, Andreas Baumgartner$^{*,1,2}$, Lukas Gubser$^{1,2}$, Christian J\"unger$^1$, Gerg{\H{o}} F{\"u}l{\"o}p$^1$, Malin Nilsson$^1$, Francesca Rossi$^3$, Valentina Zannier$^4$, Lucia Sorba$^4$, Christian Sch\"onenberger$^{1,2}$}

\address{$^1$ Department of Physics, University of Basel, Klingelbergstrasse 82, CH-4056 Basel, Switzerland}
\address{$^2$ Swiss Nanoscience Institute, University of Basel, Switzerland}
\address{$^3$ IMEM—CNR, Parco Area delle Scienze 37/A, I-43124 Parma, Italy}
\address{$^4$ NEST, Istituto Nanoscienze - CNR and Scuola Normale Superiore, Piazza San Silvestro 12, 56127 Pisa, Italy}

\date{\today}

\begin{abstract}
We present a comprehensive electrical characterization of an InAs/InP nanowire heterostructure, comprising two InP barriers forming a quantum dot (QD), two adjacent lead segments (LSs) and two metallic contacts, and demonstrate how to extract valuable quantitative information of the QD. The QD shows very regular Coulomb blockade (CB) resonances over a large gate voltage range. By analyzing the resonance line shapes, we map the evolution of the tunnel couplings from the few to the many electron regime, with electrically tunable tunnel couplings from $<$\SI{1}{\micro\electronvolt} to $>$\SI{600}{\micro\electronvolt}, and a transition from the temperature to the lifetime broadened regime. The InP segments form tunnel barriers with almost fully symmetric tunnel couplings and a barrier height of $\sim$\,$350$\,meV. All of these findings can be understood in great detail based on the deterministic material composition and geometry. Our results demonstrate that integrated InAs/InP QDs provide a promising platform for electron tunneling spectroscopy in InAs nanowires, which can readily be contacted by a variety of superconducting materials to investigate subgap states in proximitized NW regions, or be used to characterize thermoelectric nanoscale devices in the quantum regime.
\\


\end{abstract}

\maketitle
\ioptwocol
\section{Introduction}

Semiconducting nanowires (NWs), such as InAs or InSb NWs, have recently attracted significant attention, for example as building blocks in topological quantum computation \cite{Alicea2011}, sources of entangled electrons \cite{Hofstetter2009,Fueloep2015}, spintronics \cite{Nadj-Perge2010}, or thermoelectrics \cite{Karg2014}. Fundamental unique properties of these systems are their strong spin-orbit interaction \cite{Kosaka2001,Fasth2007}, large and tunable g-factors \cite{Bjoerk2005,Hollosy2013}, and are an excellent thermoelectric figure of merit (ZT) \cite{Prete2019}.  These properties make them promising material platforms for various scalable electronic devices, as well as for investigating fundamental physics on the nanometer scale. Significant progress has been made in the synthesis and bandstructure engineering of III-V semiconductors. For example, quantum dots (QDs) can be embedded into radial \cite{Jiang2007,Nilsson2016a} and axial \cite{Bjoerk2002,Dick2010,Zannier2019} NW heterostructures, directly grown complex multiple NW geometries such as crosses \cite{Gooth2017,Plissard2013,Krizek2017} or networks \cite{Gazibegovic2017} have become feasible, as well as in situ grown expitaxial superconducting shells \cite{Krogstrup2015} for superconducting hybrid devices \cite{Deng2018,Vaitiekenas2018}.

Several theoretical proposals suggest using \\semiconductor-superconductor nanowire hybrid systems to artificially create exotic quantum states of matter, such as Majorana bound states \cite{Chevallier2018,Gharavi2016,Deng2016}, and to read out qubit states \cite{Plugge2017,Leijnse2011}. NW heterostructures with in situ grown tunnel barriers are a promising platform to bring such experiments to the next level of control and to a quantitative understanding. To obtain a reliable spectroscopic tool, QDs with systematically tunable characteristics are essential. In contrast to electrostatic gating \cite{Heedt2015}, this can be achieved using in situ grown tunnel barriers in InAs NWs, either by modifying the crystal phase \cite{Nilsson2016} or by introducing a larger band gap material such as InP \cite{Bjoerk2004}. Crystal-phase defined double barriers in InAs NWs result in stable and well controllable QDs \cite{Nilsson2016}, and were recently used to probe the evolution of the superconducting proximity gap in an adjacent NW segment \cite{Juenger2019}. However, the relatively low and long tunnel barriers \cite{Chen2017,Nilsson2016} limit the spectroscopic range, while the large carrier concentrations in the zinc-blende sections make studies of few mode quantum systems challenging.

Here, we use InAs/InP NW heterostructure QDs, where a QD is formed between two InP segments in a wurtzite InAs NW. These in situ grown tunnel barriers result in a strong confinement due to a large conduction band edge offset of $V_0\approx400\,{\rm meV}-600\,$meV between the InAs and the InP segments \cite{Bjoerk2002a,Niquet2008}. Similar heterostructures have been used previously to investigate single \cite{Bjoerk2004,Romeo2012,Cornia2019} and double \cite{Fuhrer2007,Rossella2014} QD physics, as well as thermoelectric transport \cite{Prete2019}.


We present an in-depth analysis of an InAs/InP heterostructure QD demonstrating their exceptional long term stability and broad electrical tunability. We report a detailed and comprehensive characterization of the InP tunnel barriers and the resulting Coulomb blockade (CB) resonance lineshapes, which can be crucial, for example, to distinguish different single electron \cite{Lindemann2002} or superconducting subgap transport processes \cite{Fueloep2015,Gramich2015}. We show that the in situ grown InP barriers result in highly predictable, electrically tunable, and symmetric QDs with level broadenings that are small enough for high resolution spectroscopy of subgap states in hybrid systems, demonstrated by distinct  spectral features in the lead segments.

\section{Results}
\subsection*{Device Fabrication}

\begin{figure}[!h]
\includegraphics{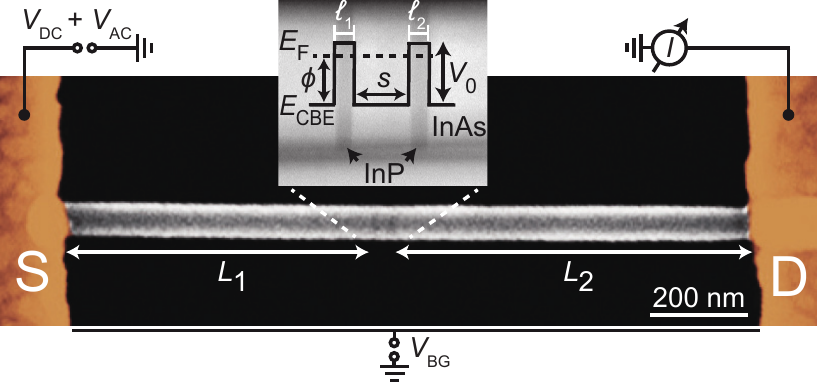}
\caption{\label{Figure1} False colored scanning electron micrograph of a typical device consisting of an InAs nanowire with two in situ grown InP tunnel barriers of length $\ell_1,\ell_2 \approx $ \SI{5.5}{\nano\meter}.  The nanowire segments between the contacts (S/D) and the quantum dot (QD) are referred to as the lead segments $LS_{1/2}$ of lengths $L_{1/2}$. A QD of length $s \approx $ \SI{19}{\nano\meter} forms between the two tunnel barriers due to the conduction band offset, $V_0$, between InAs and InP.  $V_{\rm BG}$ is the global backgate voltage that simultaneously tunes the QD and the LSs. The inset shows a transmission electron microscopy image of two InP segments pointed out by two black arrows and an energy diagram of the gate tunable conduction band edge, $E_{\rm CBE}$. The Fermi energy, $E_{\rm F}$, is indicated by a dashed line and the difference between $E_{\rm F}$ and $E_{\rm CBE}$ is defined as $\phi$.}
\end{figure}

The InAs/InP heterostructure NWs were grown by gold assisted chemical beam epitaxy \cite{Zannier2019} and have a diameter of $50 \pm 5$\,nm, depending on the size of the gold seed particle. The QD is formed on an InAs segment of length $s \approx $ \SI{19}{\nano\meter}, bounded by two InP barriers of width $\ell_1,\ell_2 \approx $ \SI{5.5}{\nano\meter}, as shown in Figure\,\ref{Figure1}. The dimensions of the InP barriers were determined in a transmission electron microscopy (TEM) analysis of NWs of the same growth. The device was fabricated on a degenerately p-doped silicon substrate acting as a global back gate with a \SI{400}{\nano\meter} thick $\text{SiO}_2$ capping layer. The electrical contacts to the NW are made of titanium/gold films with a thickness of  \SI{5}{\nano\meter}/\SI{65}{\nano\meter}. Before evaporating the contact material, the native oxide of the NWs is etched with an $\rm (NH_4)_2S_x:H_2O$ solution \cite{Suyatin2007}. A false color scanning electron microscopy (SEM) image of a typical device is shown in Figure\,\ref{Figure1}. In contrast to crystal-phase defined InAs QDs, the two InP segments can be imaged directly by standard SEM techniques with an in-lens detector \cite{Nilsson2016}.

We explicitly refer to the regions of bare InAs between the QD and the source or drain contact as the lead segments (LSs). The lead segment $LS_1$ between the QD and the source contact is $L_1 \approx $ \SI{350}{\nano\meter} long, while the lead segment $LS_2$ between the QD and the drain contact is $L_2 \approx $ \SI{600}{\nano\meter} long. The QD and the LSs are tuned simultaneously by the back gate voltage $V_{\rm BG}$, which shifts the conduction band edge, $E_{\rm CBE}(V_{\rm BG})$, relative to the Fermi energy, $E_{\rm F}$, to higher or lower values. For later, we use ${\phi(V_{\rm BG}) = E_{\rm F} - E_{\rm CBE}(V_{\rm BG})}$.

The inset of Figure\,\ref{Figure1} shows a TEM image of the epitaxially defined QD region in a similar NW. The two InP segments, indicated by black arrows, act as tunnel barriers  with a rectangular potential profile for electrons due to the atomically sharp transitions in the material composition. The barrier height depends the band gap discontinuities and residual strain. For our NW geometry, the barrier height is predicted to be $V_0 \approx 400$\,meV \cite{Niquet2008}, providing a strong confinement to the electrons in the axial direction.



All measurements were performed in a dilution refrigerator with a base temperature of \SI{\sim 30}{\milli\kelvin}. We apply a DC voltage to the source electrode to correct for small offsets ($V_{\rm DC} \approx$ \SI{50}{\micro\volt}) and superimpose an AC voltage of typically \SI{1}{\micro\volt_{\rm rms}} for lock-in detection, while the drain electrode is grounded and used for the current ($I$) measurement. The differential conductance $dI/dV_{\rm SD} = I_{\rm AC}/V_{\rm AC}$ was measured using standard lock-in techniques.

\begin{figure*}
\includegraphics{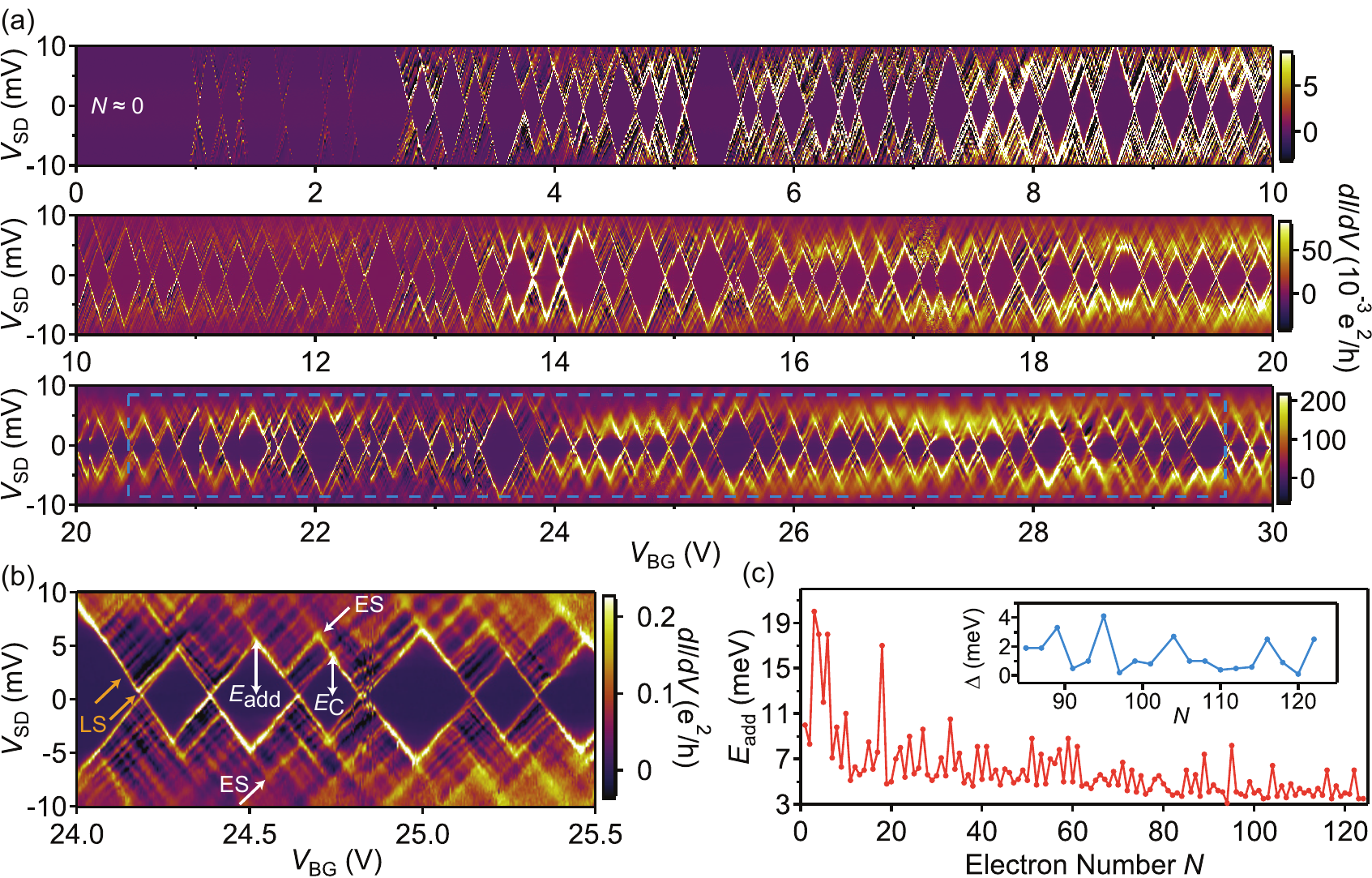}
\caption{\label{Figure2} Differential conductance, $dI/dV_{\rm SD}$, as a function of the bias, $V_{\rm SD}$, and the back gate voltage, $V_{BG}$. (a) Regular and stable Coulomb diamonds over a gate range of \SI{30}{\volt}, ranging from near depletion with an electron population of approximately zero ($N \approx 0$) to $N = 124$. (b) High resolution Coulomb diamonds, where excited state (ES) resonances of the QD and the resonances due to a modulation in the density of states in the semiconducting lead segments (LSs) are pointed out by white and orange arrows, respectively. (c) The addition energy, $E_{\rm add}$, as a function of the number of electrons on the quantum dot, $N$. The inset shows the difference of single particle energy spacing, $\Delta$, as a function of $N$ in the many electron regime extracted from the region indicated by the blue box in (a).}
\end{figure*} 

\subsection*{Characterization of the Quantum Dot}  

Figure\,\ref{Figure2}(a) shows $dI/dV_{\rm SD}$ as a function of $V_{\text{SD}}$ and $V_{\text{BG}}$. We observe regular, stable, and reproducible Coulomb diamonds (CDs) over the large backgate range of \SI{30}{\volt}, corresponding to the addition of $\sim$\,$124$ electrons. At $V_{\text{BG}} \approx $ \SI{1}{\volt}, the number of electrons on the QD is close to zero, i.e. $N \approx 0$. By increasing $V_{\text{BG}}$, electrons are added to the QD sequentially, which brings the QD into the many electron regime. For the measurement sequence shown in Figure\,\ref{Figure2}(a), the maximum number of electrons on the QD is ${N \approx 124}$. When increasing $V_{\rm BG}$ beyond \SI{40}{\volt}, thermal activation of carriers across the tunnel barriers begins to considerably contribute to the transport.

According to the constant interaction model \cite{Kouwenhoven2001}, assuming two-fold spin degenerate orbitals, the energy required to add an electron to a QD with an even electron configuration is given by the addition energy $E_{\rm add} = E_{\rm C} + \Delta$, with $E_C = e^2/C_\Sigma$ the charging energy, the total capacitance $C_\Sigma$ of the QD, and $\Delta$ the single particle energy spacing. To add a second electron to the same QD orbital requires $E_{\rm add} = E_{\rm C}$. This gives rise to an alternating even-odd pattern of large and small CDs, characteristic for spin-degenerate QD states, and allows one to extract the corresponding energy scales. A region of the CDs shown in Figure\,\ref{Figure2}(b) exhibits a clear even-odd pattern with $E_{\rm add} = $ \SI{5.5}{\milli\electronvolt} and $E_{\rm C} = $ \SI{4.2}{\milli\electronvolt}, as indicated by the white arrows. From the difference between $E_{\rm add}$ and $E_{\rm C}$, we find $\Delta = $ \SI{1.3}{\milli\electronvolt}, consistent with $\Delta\approx 1.6\,$meV from the corresponding excited state (ES) resonances outside the CDs, pointed out by white arrows.

In Figure\,\ref{Figure2}(c) we plot $E_{\rm add}$ as a function of $N$ for the full $V_{\rm BG}$ range of Figure\,\ref{Figure2}(a). We find an overall decrease in $E_{\rm add}(N)$ for increasing $N$ due to changes in the QD capacitance by electron-electron interactions \cite{Hirose1999}. From the very regular even-odd pattern in the gate range indicated by the blue box in Figure\,2(a), we extract $\Delta(N)$, as shown in the inset of Figure\,\ref{Figure2}(c).  We find that $\Delta$ strongly scatters and assumes values in between $0.2$\nobreakspace\,meV and $4$\nobreakspace\,meV, suggesting that only single levels contribute to the transport.


In addition to the QD excited state resonances, we find several other features outside of the CDs that cannot be attributed to the energy spectrum of the QD. For example, the resonances indicated by orange arrows in Figure\,\ref{Figure2}(b) are due to a non-constant DOS in the LSs, forming as standing waves in the LSs. Since these waves are strongly reflected at the InP barrier, the widths of these states are determined mostly by the coupling to the source and drain contacts, respectively. In addition, we find negative differential conductance (NDC) throughout the entire gate range, which we attribute to the simutaneous tuning of the QD and the LSs with different lever arms. We note that the NDC is more prominent in the few electron regime where the carrier concentration is low. The NDC supports our notion that the DOS in the LSs is not constant, which is typical for NW QD devices with semiconductor leads \cite{Bjoerk2004}.

\subsection*{Resonance Line Shapes}
In this section, we extract the total tunnel coupling of the QD, $\Gamma$, and the electron temperature in the LSs, $T$, from the Coulomb blockade (CB) resonances. The total tunnel coupling, $\Gamma = \Gamma_1 + \Gamma_2$, is given by the individual couplings to the source and drain leads, $\Gamma_1$ and $\Gamma_2$. In the case of an ideal measurement setup, the line shape only depends on $\Gamma$, $T$ \cite{Beenakker1991}, and the asymmetry $A = \Gamma_1 / \Gamma_2 \geq 1$ \cite{Stone1985,Buettiker1986}. However, there are also extrinsic broadening mechanisms, such as noise in the source and drain contacts, and on the gate, as well as the applied AC voltage. 


For our analysis, we assume that only a single QD level contributes to the transport, i.e. $eV_{\rm AC},\Gamma, 4k_BT \ll \Delta$,\footnote{$4k_BT$ is the 10\%-90\% width of the Fermi-Dirac distribution} and  account for the three main broadening contributions: $V_{\rm AC}$, $\Gamma$, and $T$. $V_{\rm AC}$ limits the smallest width of the line shape that can be reliably extracted, therefore $eV_{\rm AC}$ should be chosen such that $eV_{\rm AC} < 4k_BT,\Gamma$. By tuning $E_{\rm CBE}$ with $V_{\rm BG}$, we can access three different regimes: thermally broadened (${\Gamma,eV_{\rm AC} \ll 4k_BT}$), lifetime broadened (${4k_BT,eV_{\rm AC} \ll \Gamma}$), or a combination of both (${eV_{\rm AC} \ll \Gamma \approx 4k_BT}$). These three regimes are summarized in Figures\,\ref{Figure3}(a), (c), and (e), where the lifetime broadening is indicated by the width of the blue QD levels and the thermal broadening by the width of the orange Fermi-Dirac distribution in the LSs.

We model the line shape of the CB resonances with the assumption that the DOS in the LSs is constant and discuss effects due to a non-constant DOS later. For a single energy level, the line shape of a conductance resonance is described by a resonant tunneling model \cite{Beenakker1991,Ihn2009,Foxman1993}:
\begin{eqnarray}\label{eq:conductance}
I = g\frac{e}{h} \int T_{QD}(E)[f_S\left(E\right) - f_D\left(E\right)] dE,
\end{eqnarray}
where $g=1$ is the number of independent parallel transport channels, $T_{QD}(E) = (\Gamma_1\Gamma_2)/(\Delta E ^2 + \Gamma^2/4)$ the Breit-Wigner (BW) transmission function \cite{Ihn2009} with $\Delta E = E - E_0$ the detuning from the CB resonance centered at $E_0$, and ${f_{\rm S/D}(E) = 1/(1+\exp((E+eV_{\rm S/D})/k_BT))}$ are the Fermi-Dirac distributions in the LSs. $dI/dV$ is calculated numerically. The contribution of $V_{\rm AC}$ is accounted for by evaluating Equation\,\ref{eq:conductance} for a sinusoidal $V_{\rm S}$ that also electrically gates the QD. If not chosen properly, $V_{\rm AC}$ can mask the "true" resonance and the measured resonance width is then given by $V_{AC}$. 


In the regime where the broadening is mainly due to temperature, ${\Gamma \ll 4k_BT \ll \Delta}$, Equation\,\ref{eq:conductance} reduces to ${G/G_{\rm max}=\cosh^{-2}\left(\Delta E/2k_BT\right)}$, where ${G_{\rm max}= e^2/h\cdot\pi/(2k_BT)\cdot(\Gamma_1\Gamma_2/\Gamma)}$ \cite{Beenakker1991}. In this limit, $T$ can be extracted from the full-width at half maximum (FWHM) of the resonance by ${\textrm{FWHM} \approx 3.5k_BT}$. 

In the limit, where the broadening is mainly due to the electron lifetime on the QD, ${4k_BT \ll \Gamma \ll \Delta}$, Equation\,\ref{eq:conductance} reduces to the BW formula \cite{Ihn2009} $G/G_{\rm max} = (\Gamma/2)^2/(\Delta E^2 + (\Gamma/2)^2)$ with $G_{\rm max} = e^2/h\cdot 4\Gamma_1\Gamma_2/\Gamma^2$. In this limit, $\rm FWHM =$ $\Gamma$ and $A$ determines the maximum conductance $G_{\rm max} = e^2/h \cdot 4A/(1+A)^2$.




\subsection*{Evolution of the Resonance Line Shapes}

\begin{figure}
\includegraphics{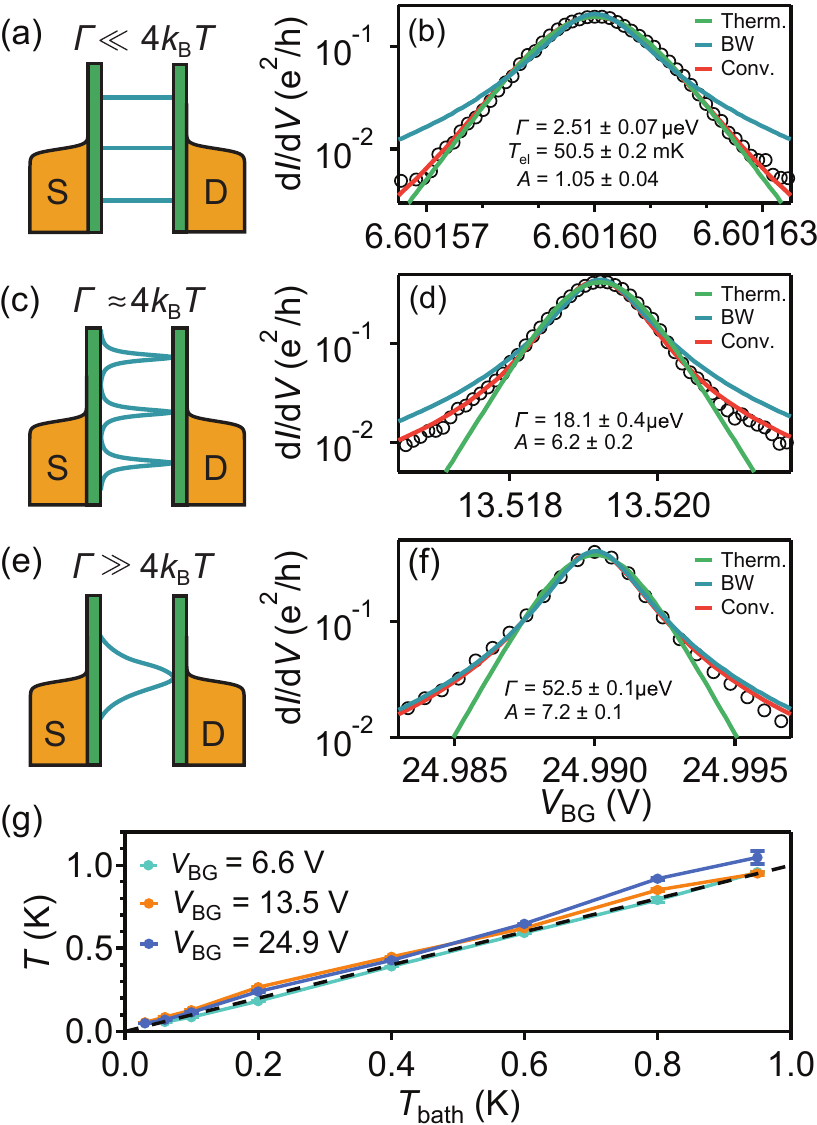}
\caption{\label{Figure3} Differential conductance $dI/dV$ as a function of $V_{\rm BG}$ for three resonances (b,d,f) in different regimes: (a) $\Gamma \ll 4k_BT$, (c) $\Gamma \approx 4k_BT$, and (e) $\Gamma \gg 4k_BT$. From (b), the electron temperature, $T = 50.5 \pm 0.2$\,mK, was extracted and was fixed in the fits of (d) and (f). (g) $T$ as a function of the  bath temperature, $T_{\rm bath}$, of the three resonances in (b), (d) and (f), respectively. $T$ remains constant for $T_{\rm bath} <$ \SI{60}{\milli\kelvin}, then increases with a slope of $1.00 \pm 0.05$, as expected for the thermally broadened regime.}
\end{figure}

We now investigate how the line shapes of the resonances evolves with $V_{\rm BG}$ and the bath temperature, $T_{\rm bath}$. Figures\,\ref{Figure3}(b),(d), and (f) show high resolution CB resonance measurements in the three broadening regimes. To show the evolution of $\Gamma$, each of the three CB resonances was fit with the expressions for a thermal, BW, and the convolution line shape, described by Equation\,\ref{eq:conductance}. From the convolution fit, we extract $\Gamma_1$, $\Gamma_2$, $T$, and their corresponding standard error of the individual fits, shown as error bars in Figure\,\ref{Figure3}(g) and \ref{Figure4}.\footnote{This error bar does not account for potential experimental errors in consecutive experiments.}

Figure\,\ref{Figure3}(b) shows a CB resonance near depletion ($N \approx 10$) at $V_{\rm BG} = $ \SI{6.6}{\volt}, measured with ${V_{\rm AC} = \SI{0.1}{\micro\volt}}$. The convolution line shape agrees very well with  the experiment, as does the pure thermal broadening line shape, but not the BW line shape. In this regime, the conduction band edge of the LSs is near the Fermi level (${\phi \ll V_0}$) and the electrons are strongly confined by the large tunnel barriers, such that the width of the Coulomb resonance is mostly determined by the electron temperature and not by the QD lifetime. Only in this regime, we can accurately determine the electron temperature of the LSs. From the convolution fit, the extracted total tunnel coupling, asymmetry, and electron temperature are ${\Gamma = 2.51 \pm 0.07}$\,\SI{}{\micro\electronvolt}, $A = 1.05 \pm 0.04$, and ${T = 50.5 \pm 0.2}$\,mK, respectively.  We see that $T$ is somewhat higher than the bath temperature (${T_{\rm bath} = 30}$\,mK), probably due to noise and radiation due to insufficient filtering. Since $T$ is not expected to change with $V_{\rm BG}$, we set $T = 50.5$\,mK for the following analysis of data at the same $T_{\rm bath}$.

For the resonance at $V_{\rm BG} =$  \SI{13.5}{\volt} ($N \approx 50$) a transition from the thermally to the lifetime broadened regime begins. As shown in Figures\,\ref{Figure3}(c) and (d), only the convolution line shape fits the data well. From the convolution fit, with $T = 50.5$\,mK and $V_{\rm AC} = $ \SI{0.25}{\micro\volt} fixed, $\Gamma = 18.1 \pm 0.4$\,\SI{}{\micro\electronvolt} and $A = 6.2 \pm 0.2$ were extracted from the fit. Therefore, this resonance is in the regime where the lifetime and thermal broadening contributes equally significantly with $\Gamma \approx 4k_BT$.

By increasing the gate voltage further, the CB resonances transition into the lifetime broadened regime with $\Gamma \gg 4k_BT$. This can be seen for the resonance in Figure\,\ref{Figure3}(f) at $V_{\rm BG} = $ \SI{24.99}{\volt} ($N \approx 100$), where the data agrees very well with the convolution fit, as well as with the BW fit, with $T = 50.5$\,mK and $V_{\rm AC} = $ \SI{1}{\micro\volt} fixed. From the convolution fit, we extract $\Gamma = 52.5 \pm 0.1$\,\SI{}{\micro\electronvolt} and $A = 7.2 \pm 0.1$, which shows that the resonance is mostly lifetime broadened ($\Gamma \gg 4k_BT$).

For each of the three resonances, the temperature dependence of the CB resonances was investigated, as shown in Figure\,\ref{Figure3}(g). We used the convolution fit with $\Gamma_1$ and $\Gamma_2$ fixed at the values determined at $T_{\rm bath} = 30$\,mK, to extract $T$ for a series of different $T_{\text{bath}}$.  For low $T_{\text{bath}}$, the CB resonances are either thermally broadened for $\Gamma \ll 4k_BT$, lifetime broadened for $4k_BT \ll \Gamma$, or a combination of the two for $\Gamma \approx 4k_BT$, as discussed in the previous section. For bath temperatures between \SI{30}{\milli\kelvin} and \SI{60}{\milli\kelvin}, the extracted $T$ remains constant. As we increase $T_{\text{bath}}$ beyond \SI{60}{\milli\kelvin}, $T$ for two resonances (cyan and orange) increases with a slope of $1.00 \pm 0.05$, in agreement with the thermally broadened regime. This is indicated by the dashed black line with a slope of $1$. However, for the blue resonance, which is mostly lifetime broadened, the slope is $1.2 \pm 0.1$, likely due to the resonance not fully transitioning into the temperature broadened regime. These experiments show that the electron and phonon system equilibrate at \SI{\sim 100}{\milli\kelvin} and that InAs/InP heterostructure QDs can be used as in-situ thermometers. In  contrast to typical Coulomb blockade thermometers \cite{Pekola1994,Palma2017}, integrated QDs form an integral part of the device, which does not require thermal coupling to a separate device.

\subsection*{Properties of the Tunnel Barriers}
\begin{figure}
\includegraphics{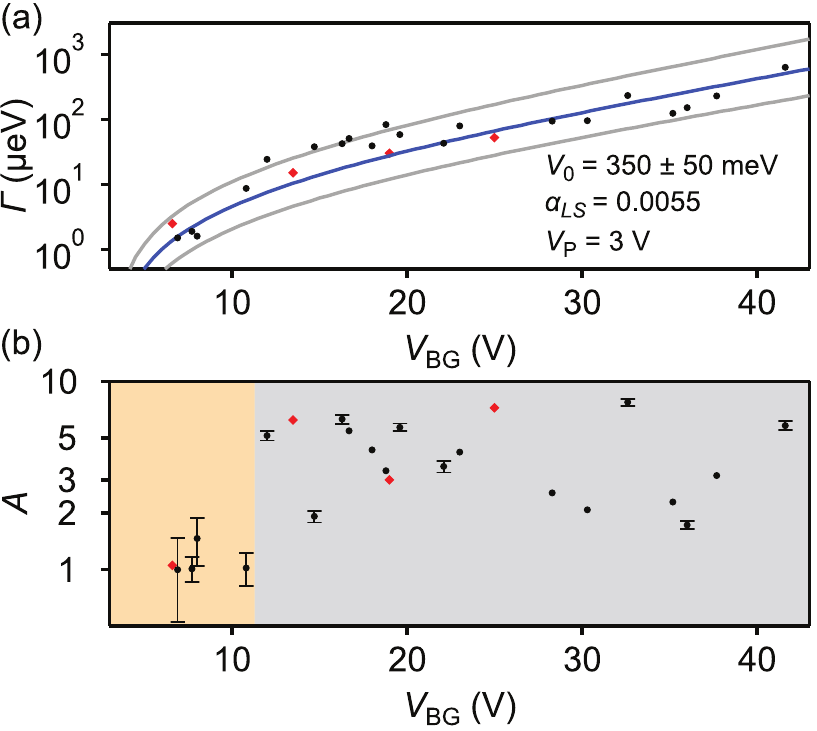}
\caption{\label{Figure4} (a) Total tunnel coupling $\Gamma$ as a function of the back gate voltage, $V_{\rm BG}$. $\Gamma$ systematically increases  with $V_{\rm BG}$ in agreement with a double barrier model described in the main text. We estimate a conduction band edge offset, $V_0$, between InAs and InP of $350 \pm 50 \,$meV (solid blue line), while the upper and lower solid gray lines are for $V_0 = 400\,$meV and $V_0 = 300\,$meV, respectively. (b) Asymmetry, $A = \Gamma_1 / \Gamma_2$, as a function of $V_{\rm BG}$. For low $V_{\rm BG}$, $A \approx 1$, while for larger $V_{\rm BG}$ $A$ scatters between $1$ and $8$. The red data points were extracted from the CB resonances in Figures\,\ref{Figure3}(b), (d), and (f) measured with a higher $V_{\rm BG}$ resolution than the black data points. Error bars smaller than the symbol size are not shown.}
\end{figure}

By investigating the functional dependence of the total tunnel coupling $\Gamma$ and the asymmetry $A$ on $V_{\rm BG}$, we estimate the height and symmetry of the tunnel barriers formed by the InP segments. By fitting the CB resonances with Equation \ref{eq:conductance} and using the previously determined $T = 50.5$\,mK, we extract $\Gamma$ and $A$ as a function of $V_{\rm BG}$, as shown in Figures\,\ref{Figure4}(a) and (b), respectively.  The red data points correspond to the CB resonances from Figures\,\ref{Figure3}(b),(d),and (f) measured with a high resolution in $V_{\rm BG}$, while the black data points stem from resonances selected from a large gate sweep ($\Delta N \approx 150$) over \SI{50}{\volt} measured with a lower resolution.


$\Gamma(V_{\rm BG})$ is plotted in Figure\,\ref{Figure4}(a) and shows a systematic increase of $\Gamma$ with increasing $V_{\rm BG}$. Close to full depletion, we find a tunnel coupling of ${\Gamma \approx \SI{1}{\micro\electronvolt}}$, which increases up to \SI{\sim 600}{\micro\electronvolt} for ${V_{\rm BG} = \SI{42}{\volt}}$. Comparing the dependence of $\Gamma(V_{\rm BG})$ to a resonant tunneling model allows us to estimate $V_0$. For this, we assume that an electron  bounces back and fourth in the InAs segment between the two InP barriers at an attempt frequency $\nu$ and escapes through either of the barriers with a probability given by the rectangular tunnel barriers. Consequently, the total tunnel coupling $\Gamma$ as a function of $V_{\rm BG}$ can be described by \cite{Ihn2009_347}:
\begin{equation}\label{eq:tunneling}
\Gamma(V_{\rm BG}) = 2\hbar\nu \left(1 + \frac{V_0^2 \sinh^2(\kappa(\phi) l)}{4\phi(V_0 - \phi)}\right)^{-1},
\end{equation}
with $\kappa(\phi) = \sqrt{2m_{\rm InP}(V_0 - \phi(V_{\rm BG}))/\hbar^2}$, ${m_{\rm InP} = 0.08m_e}$ the effective electron mass in the InP segments \cite{Kim2009}, $\phi(V_{\rm BG}) = e\alpha_{\rm LS}(V_{\rm BG} - V_{\rm P})$, $\alpha_{\rm LS}$ the lever arm of the LSs, $V_{\rm P}$ the pinch-off gate voltage, $\nu = v_{\rm F}/2s$ the attempt frequency with Fermi velocity $v_{\rm F} = \sqrt{\frac{2\phi}{m_{\rm InAs}}}$, and $m_{\rm InAs} = 0.04m_e$ the effective electron mass in wurzite InAs \cite{De2010}. The values for the length of the InP segments and the QD were taken from the TEM analysis, with $\ell = \ell_{1/2} =$ \SI{5.5}{\nano\meter} and $s =$ \SI{19.0}{\nano\meter}, respectively.

From the best fit of Equation \ref{eq:tunneling} to $\Gamma(V_{\rm BG})$ (solid blue), we obtain the free parameters ${V_0 = 350 \pm 50}$\,meV, $V_{\rm P} = 3 \rm V$, and $\alpha_{\rm LS} = 0.0053$. $V_0$ is in good agreement with the calculated literature value of $V_0 = 400\,$meV for strained InP in InAs NWs with our geometry \cite{Niquet2008}. The upper and lower solid gray lines are obtained using the same parameters and $V_0 = 300\,$meV and $V_0 = 400\,$meV, respectively. $V_{\rm P}$ agrees very well with the first CB resonances and $\alpha_{\rm LS}$ is $4.5$ times smaller than the lever arm to the QD, in qualitative agreement with the LSs being longer than the QD.


Next, we investigate the asymmetry $A$ as a function of $V_{\rm BG}$ in Figure\,\ref{Figure4}(b). The values of $A$ scatter seemingly random between $1$ and $8$ for ${V_{\rm BG} > \SI{10.5}{\volt}}$. However, for $V_{\rm BG} < $ \SI{10.5}{\volt}, $A \approx 1$ is constant, indicating highly symmetrical tunnel barriers. These characteristics of $A$ can be understood qualitatively by the following argument. The modulation of the DOS in the confined LSs is determined by the single particle level spacing in the LSs, $\Delta_{\rm LS}$, and the broadening of the energy levels in the LSs, $\Gamma_{\rm LS}$. At $E_{\rm F}$, ${\Delta_{\rm LS} = \pi \hbar v_{\rm F}/L_{1/2}}$ for a parabolic dispersion relation and thus $\Delta_{\rm LS} \sim \Delta /10$. In addition, the strong coupling between the LSs and the source or the drain contact gives rise to a larger $\Gamma_{\rm LS}$ than for the QD. With increasing $V_{\rm BG}$, $v_{\rm F}$ also increases and we suspect that for $V_{\rm BG} > $ \SI{10.5}{\volt}, $\Delta_{\rm LS} > \Gamma_{\rm LS}$, leading to a weaker overlap between the energy levels and thus to a stronger modulation of the DOS in the LSs. In contrast, for $V_{\rm BG} < $ \SI{10.5}{\volt}, $\Delta_{\rm LS}$ decreases and  the energy levels in the LSs overlap stronger, resulting in a weaker modulation of the DOS. Consequently, in the low gate regime, $A$ reflects the asymmetry of the tunnel barriers $A \approx 1$, which are essentially equal in length and height.

\section{Summary and Conclusion}

In summary, we present an in-depth characterization of a QD formed by InP tunnel barriers and connected to  metallic contacts via NW lead segments. For this system we demonstrate a nearly depletable QD with Coulomb diamonds that are exceptionally robust against charge rearrangements over a large gate range of \SI{30}{\volt}, corresponding to ${\sim\,124}$ electron states, and several months measurement time. By analyzing the line shapes of the CB resonances, we find a continuous transition from the lifetime to the thermally broadened regime and extract the electron temperature in the LSs. The QD shows a systematic and tunable increase in the tunnel coupling, based on which we estimate the conduction band edge offset between the InAs and the InP segments as ${V_0 = 350 \pm 50}$\,meV. The InP segments act like ideal tunnel barriers with an asymmetry of ${A = \Gamma_1 / \Gamma_2 \approx 1}$, as targeted in the crystal growth. This is found for low $V_{\rm BG}$, where the modulation of the DOS in the LSs is negligible, while at larger $V_{\rm BG}$ the transport is modulated by the NW lead states. In conclusion, we demonstrate that integrated InAs/InP quantum dots are a promising platform for quantitative in situ electron tunneling spectroscopy and thermometry for future superconducting hybrid devices and other electronic and thermoelectrical applications.

\ack
This research was supported by the Swiss National Science Foundation through a) a project grant entitled "Quantum Transport in Nanowires" granted to C.S. b) the National Centre of Competence in Research Quantum Science and Technology and c) the QuantEra project SuperTop. It has further been supported by a PhD grant from the Swiss Nanoscience Institut (SNI) and the University of Basel. This project has also received support from European Union's Horizon 2020 research and innovation programme under grant agreement No 828948, project AndQC. The authors declare no competing financial interest. All data in this publication are available in numerical form at DOI: \url{https://doi.org/10.5281/zenodo.3417090}

\section*{Corresponding Authors}
$^*$Email: frederick.thomas@unibas.ch \\
$^*$Email: andreas.baumgartner@unibas.ch

\section*{Author Contributions}
F.T. fabricated the device and performed the measurements. F.T., A.B., M.N., and G.F. analyzed the data. C.J. supported the fabrication and performed several electrical characterization measurements on different growth batches of the InAs/InP NWs. L.G. fabricated, measured, and analyzed a similar device. G.F. performed measurements for a similar device that F.T. and C.J. fabricated. F.R., V.Z., and L.S. developed the nanowire structure. C.S. and A.B. planned and designed the experiments, and participated in all discussions. All authors contributed to the manuscript.

\section*{References}

\bibliographystyle{iopart-num}
\bibliography{bibtex_database}

\providecommand{\newblock}{}
\begin{thebibliography}{10}
\expandafter\ifx\csname url\endcsname\relax
  \def\url#1{{\tt #1}}\fi
\expandafter\ifx\csname urlprefix\endcsname\relax\def\urlprefix{URL }\fi
\providecommand{\eprint}[2][]{\url{#2}}

\bibitem{Alicea2011}
Alicea J, Oreg Y, Refael G, von Oppen F and Fisher M~P~A 2011 {\em Nature
  Physics\/} {\bf 7} 412--417

\bibitem{Hofstetter2009}
Hofstetter L, Csonka S, Nyg{\aa}rd J and Schönenberger C 2009 {\em Nature\/}
  {\bf 461} 960--963

\bibitem{Fueloep2015}
F{\"u}l{\"o}p G, Dom{\'{\i}}nguez F, d'Hollosy S, Baumgartner A, Makk P, Madsen
  M, Guzenko V, Nyg{\aa}rd J, Sch{\"o}nenberger C, Yeyati A~L and Csonka S 2015
  {\em Physical Review Letters\/} {\bf 115} 227003

\bibitem{Nadj-Perge2010}
Nadj-Perge S, Frolov S~M, Bakkers E~P~A~M and Kouwenhoven L~P 2010 {\em
  Nature\/} {\bf 468} 1084--1087

\bibitem{Karg2014}
Karg S~F, Troncale V, Drechsler U, Mensch P, Kanungo P~D, Schmid H, Schmidt V,
  Gignac L, Riel H and Gotsmann B 2014 {\em Nanotechnology\/} {\bf 25} 305702

\bibitem{Kosaka2001}
Kosaka H, Kiselev A, Baron F, Kim K~W and Yablonovitch E 2001 {\em Electronics
  Letters\/} {\bf 37} 464

\bibitem{Fasth2007}
Fasth C, Fuhrer A, Samuelson L, Golovach V~N and Loss D 2007 {\em Physical
  Review Letters\/} {\bf 98} 266801

\bibitem{Bjoerk2005}
Björk M~T, Fuhrer A, Hansen A~E, Larsson M~W, Fröberg L~E and Samuelson L
  2005 {\em Physical Review B\/} {\bf 72} 201307

\bibitem{Hollosy2013}
D'Hollosy S, F\'{a}bi\'{a}n G, Baumgartner A, Nyg\r{a}rd J and Schönenberger C
  2013 {\em AIP Conf. Proc.\/}  359--360

\bibitem{Prete2019}
Prete D, Erdman P~A, Demontis V, Zannier V, Ercolani D, Sorba L, Beltram F,
  Rossella F, Taddei F and Roddaro S 2019 {\em Nano Letters\/} {\bf 19}
  3033--3039

\bibitem{Jiang2007}
Jiang X, Xiong Q, Nam S, Qian F, Li Y and Lieber C~M 2007 {\em Nano Letters\/}
  {\bf 7} 3214--3218

\bibitem{Nilsson2016a}
Nilsson M, Namazi L, Lehmann S, Leijnse M, Dick K~A and Thelander C 2016 {\em
  Physical Review B\/} {\bf 94} 115313

\bibitem{Bjoerk2002}
Bj{\"o}rk M~T, Ohlsson B~J, Sass T, Persson A~I, Thelander C, Magnusson M~H,
  Deppert K, Wallenberg L~R and Samuelson L 2002 {\em Nano Letters\/} {\bf 2}
  87--89

\bibitem{Dick2010}
Dick K~A, Thelander C, Samuelson L and Caroff P 2010 {\em Nano Letters\/} {\bf
  10} 3494--3499

\bibitem{Zannier2019}
Zannier V, Rossi F, Ercolani D and Sorba L 2019 {\em Nanotechnology\/} {\bf 30}
  094003

\bibitem{Gooth2017}
Gooth J, Borg M, Schmid H, Schaller V, Wirths S, Moselund K, Luisier M, Karg S
  and Riel H 2017 {\em Nano Letters\/} {\bf 17} 2596--2602

\bibitem{Plissard2013}
Plissard S~R, van Weperen I, Car D, Verheijen M~A, Immink G~W~G, Kammhuber J,
  Cornelissen L~J, Szombati D~B, Geresdi A, Frolov S~M, Kouwenhoven L~P and
  Bakkers E~P~A~M 2013 {\em Nature Nanotechnology\/} {\bf 8} 859--864

\bibitem{Krizek2017}
Krizek F, Kanne T, Razmadze D, Johnson E, Nyg{\aa}rd J, Marcus C~M and
  Krogstrup P 2017 {\em Nano Letters\/} {\bf 17} 6090--6096

\bibitem{Gazibegovic2017}
Gazibegovic S, Car D, Zhang H, Balk S~C, Logan J~A, de~Moor M~W~A, Cassidy M~C,
  Schmits R, Xu D, Wang G, Krogstrup P, het Veld R~L~M~O, Zuo K, Vos Y, Shen J,
  Bouman D, Shojaei B, Pennachio D, Lee J~S, van Veldhoven P~J, Koelling S,
  Verheijen M~A, Kouwenhoven L~P, Palmstr{\o}m C~J and Bakkers E~P~A~M 2017
  {\em Nature\/} {\bf 548} 434--438

\bibitem{Krogstrup2015}
Krogstrup P, Ziino N~L~B, Chang W, Albrecht S~M, Madsen M~H, Johnson E,
  Nyg{\aa}rd J, Marcus C~M and Jespersen T~S 2015 {\em Nature Materials\/} {\bf
  14} 400--406

\bibitem{Deng2018}
Deng M~T, Vaitiek{\.{e}}nas S, Prada E, San-Jose P, Nyg{\aa}rd J, Krogstrup P,
  Aguado R and Marcus C~M 2018 {\em Physical Review B\/} {\bf 98} 085125

\bibitem{Vaitiekenas2018}
Vaitiek{\.{e}}nas S, Deng M~T, Nyg{\aa}rd J, Krogstrup P and Marcus C 2018 {\em
  Physical Review Letters\/} {\bf 121} 037703

\bibitem{Chevallier2018}
Chevallier D, Szumniak P, Hoffman S, Loss D and Klinovaja J 2018 {\em Physical
  Review B\/} {\bf 97} 045404

\bibitem{Gharavi2016}
Gharavi K, Hoving D and Baugh J 2016 {\em Physical Review B\/} {\bf 94} 155417

\bibitem{Deng2016}
Deng M~T, Vaitiek{\.{e}}nas S, Hansen E~B, Danon J, Leijnse M, Flensberg K,
  Nyg{\aa}rd J, Krogstrup P and Marcus C~M 2016 {\em Science\/} {\bf 354}
  1557--1562

\bibitem{Plugge2017}
Plugge S, Rasmussen A, Egger R and Flensberg K 2017 {\em New Journal of
  Physics\/} {\bf 19} 012001

\bibitem{Leijnse2011}
Leijnse M and Flensberg K 2011 {\em Physical Review B\/} {\bf 84} 140501

\bibitem{Heedt2015}
Heedt S, Otto I, Sladek K, Hardtdegen H, Schubert J, Demarina N, Lüth H,
  Grützmacher D and Schäpers T 2015 {\em Nanoscale\/} {\bf 7} 18188--18197

\bibitem{Nilsson2016}
Nilsson M, Namazi L, Lehmann S, Leijnse M, Dick K~A and Thelander C 2016 {\em
  Physical Review B\/} {\bf 93} 195422

\bibitem{Bjoerk2004}
Bj{\"o}rk M~T, Thelander C, Hansen A~E, Jensen L~E, Larsson M~W, Wallenberg L~R
  and Samuelson L 2004 {\em Nano Letters\/} {\bf 4} 1621--1625

\bibitem{Juenger2019}
Jünger C, Baumgartner A, Delagrange R, Chevallier D, Lehmann S, Nilsson M,
  Dick K~A, Thelander C and Schönenberger C 2019 {\em Communications
  Physics\/} {\bf 2}

\bibitem{Chen2017}
Chen I~J, Lehmann S, Nilsson M, Kivisaari P, Linke H, Dick K~A and Thelander C
  2017 {\em Nano Letters\/} {\bf 17} 902--908

\bibitem{Bjoerk2002a}
Björk M~T, Ohlsson B~J, Sass T, Persson A~I, Thelander C, Magnusson M~H,
  Deppert K, Wallenberg L~R and Samuelson L 2002 {\em Applied Physics
  Letters\/} {\bf 80} 1058--1060

\bibitem{Niquet2008}
Niquet Y~M and Mojica D~C 2008 {\em Physical Review B\/} {\bf 77} 115316

\bibitem{Romeo2012}
Romeo L, Roddaro S, Pitanti A, Ercolani D, Sorba L and Beltram F 2012 {\em Nano
  Letters\/} {\bf 12} 4490--4494

\bibitem{Cornia2019}
Cornia S, Rossella F, Demontis V, Zannier V, Beltram F, Sorba L, Affronte M and
  Ghirri A  (\textit{Preprint} \eprint{http://arxiv.org/abs/1907.12324v1})

\bibitem{Fuhrer2007}
Fuhrer A, Fröberg L~E, Pedersen J~N, Larsson M~W, Wacker A, Pistol M~E and
  Samuelson L 2007 {\em Nano Letters\/} {\bf 7} 243--246

\bibitem{Rossella2014}
Rossella F, Bertoni A, Ercolani D, Rontani M, Sorba L, Beltram F and Roddaro S
  2014 {\em Nature Nanotechnology\/} {\bf 9} 997--1001

\bibitem{Lindemann2002}
Lindemann S, Ihn T, Bieri S, Heinzel T, Ensslin K, Hackenbroich G, Maranowski K
  and Gossard A~C 2002 {\em Physical Review B\/} {\bf 66} 161312

\bibitem{Gramich2015}
Gramich J, Baumgartner A and Schönenberger C 2015 {\em Physical Review
  Letters\/} {\bf 115} 216801

\bibitem{Suyatin2007}
Suyatin D~B, Thelander C, Björk M~T, Maximov I and Samuelson L 2007 {\em
  Nanotechnology\/} {\bf 18} 105307

\bibitem{Kouwenhoven2001}
Kouwenhoven L~P, Austing D~G and Tarucha S 2001 {\em Reports on Progress in
  Physics\/} {\bf 64} 701--736

\bibitem{Hirose1999}
Hirose K and Wingreen N~S 1999 {\em Physical Review B\/} {\bf 59} 4604--4607

\bibitem{Beenakker1991}
Beenakker C~W~J 1991 {\em Physical Review B\/} {\bf 44} 1646--1656

\bibitem{Stone1985}
Stone A~D and Lee P~A 1985 {\em Physical Review Letters\/} {\bf 54} 1196--1199

\bibitem{Buettiker1986}
Büttiker M 1986 {\em Physical Review B\/} {\bf 33} 3020--3026

\bibitem{Ihn2009}
Ihn T 2009 {\em Semiconductor Nanostructures\/} (Oxford University Press)

\bibitem{Foxman1993}
Foxman E~B, McEuen P~L, Meirav U, Wingreen N~S, Meir Y, Belk P~A, Belk N~R,
  Kastner M~A and Wind S~J 1993 {\em Physical Review B\/} {\bf 47} 10020--10023

\bibitem{Pekola1994}
Pekola J~P, Hirvi K~P, Kauppinen J~P and Paalanen M~A 1994 {\em Physical Review
  Letters\/} {\bf 73} 2903--2906

\bibitem{Palma2017}
Palma M, Scheller C~P, Maradan D, Feshchenko A~V, Meschke M and Zumbühl D~M
  2017 {\em Applied Physics Letters\/} {\bf 111} 253105

\bibitem{Ihn2009_347}
Ihn T 2009 {\em Semiconductor Nanostructures\/} 347 (Oxford University Press)

\bibitem{Kim2009}
Kim Y~S, Hummer K and Kresse G 2009 {\em Physical Review B\/} {\bf 80} 035203

\bibitem{De2010}
De A and Pryor C~E 2010 {\em Physical Review B\/} {\bf 81} 155210

\end{thebibliography}

\end{document}